\documentclass[groupedaddress,superscriptaddress,aps,showpacs,amsmath,amssymb,floatfix, prl,twocolumn,a4]{revtex4-1}

 \usepackage{graphicx,float}
 \usepackage{subfigure}
 \usepackage{dcolumn}
\usepackage{bm}
\usepackage{mathrsfs}
\usepackage{txfonts}
\usepackage{CJK}
\usepackage{SIunits}
\usepackage{epsfig}
\usepackage{lipsum}
\usepackage{color}

\usepackage{placeins}

\newcommand{\LaplaceOpr}{\bigtriangledown^2}
\newcommand{\PDt}[1]{\frac{\partial {#1}}{\partial t}}
\newcommand{\PDtsq}[1]{\frac{\partial^2 {#1}}{\partial t^2}}

 \newenvironment{SChinese}{%
  \CJKfamily{gbsn}%
 \CJKtilde
  \CJKnospace}{}

\allowdisplaybreaks[2]

\begin{document}

\begin{CJK}{UTF8}{} 
\begin{SChinese}

\title{Giant enhancement of tunable optomechanical coupling via ultrarefractive medium}

 \author{Keyu Xia (夏可宇)}  %
 \email{keyu.xia@mq.edu.au}
 \affiliation{Centre for Engineered Quantum Systems, Department of Physics and Astronomy, Macquarie University, NSW 2109, Australia}

 \author{Jason Twamley}
 \affiliation{Centre for Engineered Quantum Systems, Department of Physics and Astronomy, Macquarie University, NSW 2109, Australia}

\date{\today}

\begin{abstract}
 Exploring the fundamental quantum behaviour of optomechanical resonators is of great interest recently but requires the realization of the strong coupling regime.
  We study the optical photon-phonon coupling of the so-called membrane in the middle (MITM) optomechanical system. 
  Using  coupled-mode theory we find that the optomechanical coupling is proportional to the electric susceptibility of the membrane. By considering the doping atoms or spins into the membrane and driving these appropriately  we induce a tunable ultra-large refractive index without absorption which enhances the optomechanical coupling. Using this we  predict an ultra-strong single-optical photon strong coupling  with large quantum cooperativity for $\text{Er}^{3+}$ dopants at low temperature, while  $\text{Cr}^{3+}$ in a Ruby membrane may display ultra-large quantum cooperativity at {\em room temperature}. Our scheme also can tune the strength of the coupling over a wide range and can also control whether the optomechanical force is attractive or repulsive. Our work opens a door for fundamental physics and applications relying on the realization of the strong coupling regime in quantum optomechanical systems. 
\end{abstract}


\maketitle

\end{SChinese}
 \end{CJK}

Optomechanical systems have made tremendous progress over the past decade. It is a promising testbed for fundamental quantum mechanics from manipulating the motional quantum ground state of mescoscopic objects \cite{GroundStCooling1,*GroundStCooling2,CPSun,*xia1,*xia2}, to non-Gaussian quantum state synthesis \cite{PhysRevLett.107.063602}, and also a platform for important applications such as squeezed light \cite{squeezing1,*squeezing2}, ultrasensitive measurement \cite{OptomMeas1}, diabolical points \cite{DiabolicalPointJason}, and microwave(mw)-to-optical quantum interface and quantum networks \cite{milburnvitali,QNet1}.

 The above schemes crucially rely on amplifying the rather weak single-photon coupling $g_\text{om}$. 
Observation of a non-Gaussian state of the mechanical resonator even requires a single-photon coupling stronger than the linewidth $\kappa$ of the cavity, i.e. the so-called single-photon strong coupling (SPSC) regime \cite{PhysRevLett.107.063602}. SPSC has only been demonstrated so far in optomechanical arrangements utilizing ultracold atomic gases \cite{ColdAtom1,*ColdAtom2}. SPSC continues to remain a very challenging regime for mescoscopic mechanical resonators. The state-of-the-art current experiments in mesocopic optomechanics typically can only achieve the small coupling-decay ratio (CDR) of $10^{-3}$ \cite{OptoRev2}. Even after amplification by a large optical coherent state $|\alpha\rangle$, the record experimental CDR for optical setups $\sim 1.5$ \cite{MVanner}. On the other hand, there is much interest in the quantum behavior of optomechanical systems which have a quantum cooperativity (QC) in excess of unity \cite{QuantCoh,OptoRev2}. In this case the light and mechanics can coherently exchange faster than the thermal decoherence of mechanical motion. Only very recently, two theoretical proposals making use of ultrasensitive superconducting circuits \cite{EnhSupercond1,EnhSupercond2} approach the SPSC regime for microwave (mw) photons, $g_\text{om}\gtrsim \kappa$, but require  cryogenic temperatures. An array of identical mechanical resonators supporting a collective oscillation is also theoretically studied for strong coupling to an optical mode \cite{EnhCollective,DiabolicalPointJason}, while a very recent theory proposal aims to achieve SPSC via an optically trapped nanodiamond \cite{Thomas}.  However, in typical setups the coupling between a single mechanical mode and optical mode is weak.
%

\begin{figure}
 \centering
\setlength{\unitlength}{1cm}
\begin{picture}(9,3.5)
\put(0,-.7){ \includegraphics[width=.95\linewidth]{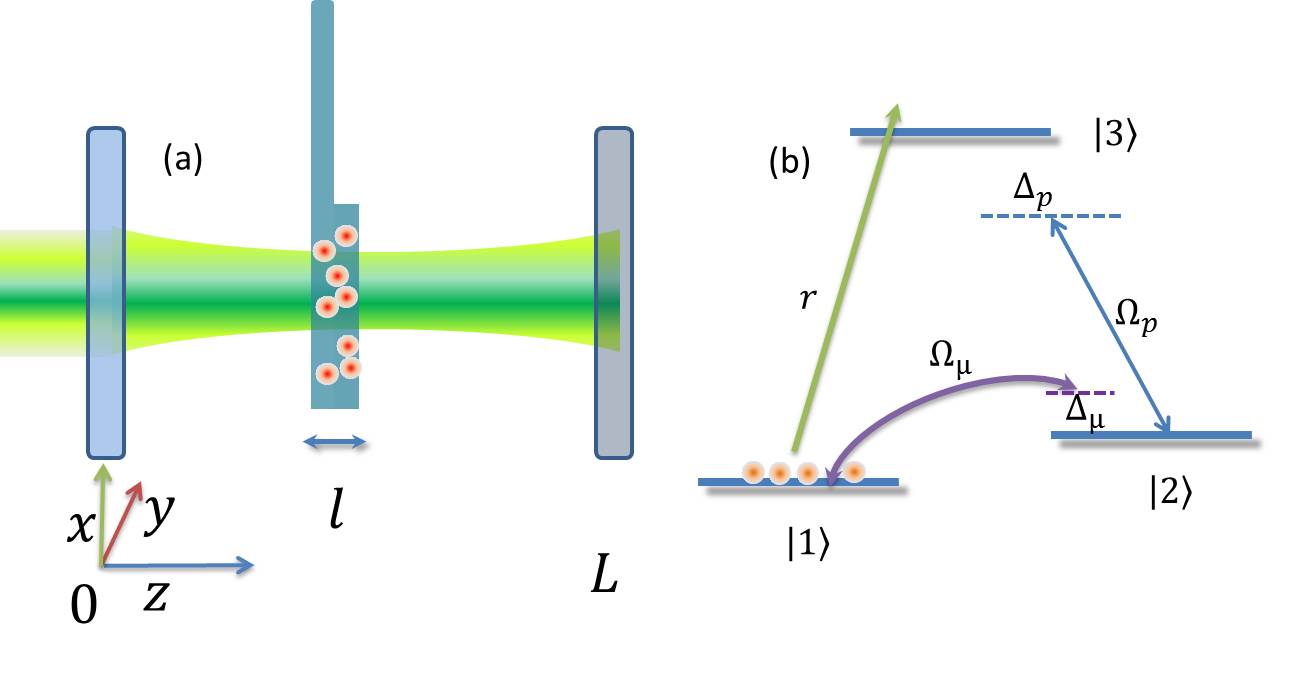}}
\end{picture}
 \caption{(Color online). (a) Schematic of the optomechanical resonator. A membrane is inserted into an optical Fabry-P\'erot cavity. An ensemble of $\Lambda$-type atomic systems  are doped in the membrane. (b) The $\Lambda$-atomic configuration is optically but incoherently pumped from its ground state $|1\rangle$ to the excited state $|3\rangle$. A coherent microwave field detuned by $\Delta_\mu$ is applied to create coherence between the doublet ground states $|1\rangle$ and $|2\rangle$. The probe (cavity mode $\hat{a}$) field detuned by $\Delta_p$ drives the transition $|2\rangle \leftrightarrow |3\rangle$.}\label{fig:system}
\end{figure}

In this letter, we propose a dynamically tunable optomechanical system, see Fig. \ref{fig:system}. We find that the optomechanical coupling is proportional to the electric susceptibility $\chi$ of the membrane. Based on this we propose a theoretical scheme  to enhance the electric susceptibility of the mechanical resonator by several orders in magnitude. We greatly enhance the optomechanical coupling  without introducing extra loss to the cavity, to achieve the SPSC regime.
%
Ultrarefractive media have been  studied in various systems \cite{UltraNTheo1,UltraNTheo2,UltraNTheo3}, and has been demonstrated in experiment \cite{UltraNExp}, but it has not been exploited for enhancing optomechanical coupling so far.

One implementation of our scheme is depicted in Fig. \ref{fig:system}(a). A membrane with thickness $l$ is inserted at equilibrium position $z_0$ in a Fabry-P\'erot(FP) cavity with length $L$ and finesse $F$, a setup well studied in experiments \cite{Membrane1}. The membrane has mass $M$ and oscillates with frequency $\Omega_m$, with zero-point motion $z_\text{zp}=\sqrt{\hbar/2M\Omega_m}$, and we represent the quantised membrane's displacement  from $z_0$ as $z_m=z_\text{zp} (\hat{b}^\dag + \hat{b})$, where $\hat{b}^\dag(\hat{b})$ are the creation(annihilation) operators of the mechanical motion.  The membrane is made from a host medium with a relative permittivity $\varepsilon_\text{h}$. An ensemble of $\Lambda$-type quantum systems is doped in the membrane to create extra susceptibility $\chi$, and whose atomic level system is schematically shown in  Fig. \ref{fig:system}(b). 
This configuration can induce a giant electric susceptibility to substantially modify the phase of a probe field passing through it. 

\emph{Mode coupling} 
 Using coupled-mode theory \cite{CMTYariv, *CMTHuang,*CMTXia1,*CMTXia2,CMTPainter1,*CMTPainter2}, we present a derivation of the optomechanical coupling strength involving many more parameters and from this we will observe the dependence on the electric susceptibility of the membrane.  We first derive the coupling between different cavity modes without the mechanical motion of the membrane. We consider the membrane in Fig. \ref{fig:system}(a) as a slab of medium with a macroscopic complex polarization $P=\varepsilon_0 (\chi+\varepsilon_h-1) E_p$, where $\varepsilon_0$ is the vacuum permittivity, and $E_p$ is the probe field input into the membrane. The overall relative permittivity is $\chi+\varepsilon_h$. We set $\xi=\xi^\prime+\xi''$, where $\xi^\prime=\text{Re}[\xi]$ and $\xi''=\text{Im}[\xi]$.
%
%
We consider now an empty optical FP cavity. The total electric field trapped in the cavity is $\vec{E}_p$, and we seperate this  into the positive and negative frequency components, $\vec{E}=\vec{E}^+ + \vec{E}^-$, $\vec{E}^\pm = e^{\mp j\omega t} \vec{E}(\bf{r})$, where $E({\bf r})$ is the E-field spatial distribution at position $\bf{r}$ in the cavity. This E-field obeys Maxwell's wave equation 
$\LaplaceOpr \vec{E} - \frac{1}{c^2} \PDtsq{\varepsilon_0 \varepsilon_r({\bf r}) \vec{E}}=0$,
where $\varepsilon_r({\bf r})$ is the overall relative permittivity of the optical cavity at position ${\bf r}$,  $c$ is the speed of light  in vacuum and  $\varepsilon_r({\bf r})=1$, in the absence of the membrane. The E-field in the cavity can be decomposed as a superposition of eigenmodes  as
\begin{equation}\label{eq:EtoEl}
 \vec{E}^+ = \sum_l \hat{a}_l(t)\, \vec{E}^{+}({\bf r},t) = \sum_l \vec{e}_l \,\hat{a}_l(t) \,e^{-j\omega_l t}\, E_l({\bf r})  \;,
\end{equation}
where $E_l({\bf r})$ is the E-field distribution for the $l$'\emph{th} eigenmode, $\omega_l$ is the resonant frequency and $\vec{e}_l$ is the polarization of this mode and $\hat{a}_l(\hat{a}^\dag_l)$ is the annihilation and creation operator of the $l$'th mode.
For each eigenmode, we have 
  $c^2\LaplaceOpr \vec{E}_l^+ - \partial^2 \varepsilon_0  \vec{E}_l^+/ \partial t^2=0$.
Substituting Eq. (\ref{eq:EtoEl}) into Maxwell's wave equation, we have
$\sum_l \hat{a}_l \LaplaceOpr \vec{E}_l^+ ({\bf r},t) - \frac{1}{c^2} \PDtsq{\hat{a}_l \varepsilon_r \vec{E}_l^+ ({\bf r},t)} = 0$.
It is also reasonable to assume that $\varepsilon_r$ is time-independent in the absence of a large time-dependent driving of the membrane. Normally, we have $\left|-2j\omega_l \PDt{\hat{a}_l} \right| \gg \left|\PDtsq{\hat{a}_l} \right| $ leading to $\PDtsq{\hat{a}_l \varepsilon_r \vec{E}_l^+ ({\bf r},t)}  =  \hat{a}_l \PDtsq{\varepsilon_r \vec{E}_l^+ ({\bf r},t)} + 2 \PDt{\hat{a}_l} \PDt{\varepsilon_r \vec{E}_l^+ ({\bf r},t)}$. Thus, Maxwell's wave equation, as a decomposition of the eigenmodes, has the form
\begin{equation}
 \sum_l \left\{2 \PDt{\varepsilon_r \vec{E}_l^+ ({\bf r}, t)} \PDt{\hat{a}_l} + \hat{a}_l \PDtsq{ (\varepsilon_h-1 +\chi)\vec{E}_l^+ ({\bf r}, t)} \right\} =0 \;.
\end{equation}
We focus on the modification of the $k$th mode by the mechanical motion but neglect the intermode couplings in this paper.
Using the orthogonality conditions 
 $\int_{V_\text{cavity}}\!\! \!\!\!\!\vec{e}_k\cdot \vec{e}_l \,E_k^*({\bf r})E_l({\bf r})d {\bf r} = \delta_{kl}$, the coupled mode equation in the absence of mechanical motion reads as $\PDt{\hat{a}_k} =  j\Delta\omega_k \hat{a}_k - \Delta\kappa_k \hat{a}_k$,
where $\Delta\omega_k$ and $\Delta\kappa_k$ corresponds to the frequency shift and linewidth change by inserting the membrane, respectively. One finds $2U_k\,\Delta\omega_k  = {\omega_k} {\varepsilon_0\int_{V_\text{obj}} \chi^\prime |E_k({\bf r})|^2 d {\bf r}}  + {\omega_k} {\varepsilon_0\int_{V_\text{obj}} (\varepsilon'_h-1) |E_k({\bf r})|^2 d {\bf r}} $  and $2U_k\, \Delta\kappa_k  = {\omega_k}{\varepsilon_0\int_{V_\text{obj}} (\varepsilon''_h+\chi'') |E_k({\bf r})|^2 d {\bf r}} $, with the volume of the membrane $V_\text{obj}$ and $U_k=\varepsilon_0 \int_{V_\text{cavity}}  |E_k({\bf r})|^2 d {\bf r}$. These formula are in agreement with those derived for toroidal cavities \cite{CMTPainter1,*CMTPainter2}.
Defining the E-field $E_k({\bf r}) = E_{k,0} e_k({\bf r})$, where the amplitude $E_{k,0} = \sqrt{\hbar\omega_k/\varepsilon_0 V_k }$, with the mode volume $V_k$, and $e_k({\bf r})$,  the spatial distribution of the mode (with $ \int_{V_\text{cavity}} |e_k({\bf r})|^2 d {\bf r} /V_k=1$), we take 
the normalized mode distribution to be
 $e_k({\bf r}) = f_k(x,y,z)\sin(k_k z)$,
where $f_k(x,y,z)$ is the transverse profile of the $k$'\emph{th} mode, varying slowly along the z-axis, and $k_k$ is the z-component of the propagation constant. Here we assume that the field propagates along the z-axis such that $k_k =2 \pi/\lambda_k$ with the wavelength $\lambda_k = 2\pi c/\omega_k$. In the FP cavity, $\sin(k_k L)=0$. In the case of $l\ll \lambda_k$, $f_k(x,y,z)\approx f_k(x,y,z_0)$ in the membrane.
We have $U_k = \frac{L}{2}\varepsilon_0 |E_{k,0}|^2 A_k$ with the effective mode cross section area $A_k = 2V_k/L$. For some position $z_0$, $A_k= \int\int |f_k(x,y,z_0)|^2 dxdy$.

Now we consider the mechanical motion of the membrane. The displacement $d_m$ of the membrane around its equilibrium position $z_0$ is dependent on the mechanical mode. The displacement can be described as 
 $d_m(x,y,z)=z_m f_m(x,y)$,
where $f_m(x,y)$ is the transverse mode distribution of the mechanical vibration and $z_m$ is the displacement amplitude.
Correspondingly, the electric field of the $k$\emph{th} cavity mode at the displaced membrane location becomes $e_k({\bf r}) = f_k(x,y,z_0+d_m) \sin{k_k (z_0+d_m)} \approx f_k(x,y,z_0) [\sin{k_kz_0} + k_k\cos{k_kz_0} f_m(x,y)z_m]$, expanding to  first order in $d_m$. 

Applying $z_m=z_\text{zp}(\hat{b}^\dag+ \hat{b})$, the optomechanical Hamiltonian for a specific cavity mode $\hat{a}$ takes the form, when $k l  \ll 1$,
\begin{equation} \label{eq:Hm}
\begin{split}
 H_m = & \left[\omega_c + \Delta\omega +i (\Delta\kappa +\kappa)\right] \hat{a}^\dag \hat{a} \;\\
 &  +  \left[g_\text{om,P}+ g_\text{om,h}\right]\hat{a}^\dag \hat{a}(\hat{b}^\dag+ \hat{b}) \;,
\end{split}
\end{equation}
with {$\kappa$ is the total decay rate of cavity without the membrane.  
\begin{subequations}\label{eq:coeff}
 \begin{align}
  \Delta\omega & \approx {\omega_cl}(\varepsilon'_h-1+\chi')\sin^2 (kz_0)/L \;,\\
  \Delta\kappa & \approx { Q_c\, \kappa l}\,(\varepsilon''_h+\chi'') \sin^2 (kz_0)/L \;,\\
  g_\text{om,P} & = \chi{\omega_c}\,  z_\text{zp} \,kl  \,\sin(2 k z_0) \mathcal{B}/L \;,\\
  g_\text{om,h} & =(\varepsilon_h-1)\,  {\omega_c}z_\text{zp}\,kl\,   \sin(2k z_0)\mathcal{B}/L \;,
 \end{align}
\end{subequations}
where $\mathcal{B} = \int\int |f_k(x,y,z_0)|^2 f_m(x,y) dxdy/A_k$ can be almost unity if $f_m(x,y)=1$ within the optical mode waist, and $Q_c=\omega_c/\kappa$ is the quality factor of the cavity.
 According to Eq. (\ref{eq:coeff}), the coupling is not only determined by $\omega_c/L$ but also proportional to the electric susceptibility of the material. Therefore, a giant susceptibility $\chi'$ without extra additional absorption, i.e. $\chi''\leq 0$, can greatly enhance the optomechanical coupling strength. 

We choose to separate the couplings into a contributions due to the host material, $g_\text{om,h}$, and the induced polarization $g_\text{om,P}$ due to the doped atomic systems.
The total optomechanical coupling is $g_\text{om}=g_{\text{om},P} + g_\text{om,h}$. We now explore how our scheme improves the of-interest coupling-decay ratio (CDR). We assume that the intrinsic loss rate of the cavity is $\kappa_i$. With critical external coupling we have $\kappa_\text{ex} = \kappa_i$ and $\kappa=\kappa_\text{ex} + \kappa_i$, and in the absence of the atomic polarization the CDR is $g_\text{om,h}/\kappa$. Considering now the induced polarization of the atoms the intrinsic decay becomes $\kappa_i^\prime = \kappa_i + \Delta\kappa$ and the total decay changes to $\kappa^\prime = 2\kappa_i + \Delta\kappa$ under the critical coupling condition. After including the change in decay rates, the improved CDR becomes
\begin{equation} \label{eq:ratio}
  \frac{g_\text{om}}{\kappa^\prime}  = \frac{\kappa\,\chi}{\kappa^\prime (\varepsilon_h-1)} \frac{g_\text{om,h}}{\kappa}\;.
\end{equation}
We are interested when, $|\chi| \gg (\varepsilon_h-1)$, yielding $G\approx \chi/(\varepsilon_h-1)$.
%

\emph{Giant susceptibility} We now present a method to induce a giant polarization without extra absorption (and potentially gain), using a $\Lambda$-type atomic level system, see Fig. \ref{fig:system}(b). The advantages are twofold: (i) the giant susceptibility $\chi$ can enhance the optomechanical coupling; (ii) any gain can reduce the decay rate of the cavity. For this purpose, a small population in the excited state and a quantum  coherence between the ground states are essential \cite{UltraNTheo3}. To proceed we assume the system is initially populated in $|1\rangle$ but is incoherently pumped to the excited state $|3\rangle$ with rate $r$. A mw field $\Omega_\mu$ is applied to create coherence between the two ground states.
The cavity mode $\hat{a}$ is modeled as a probe field, $\Omega_p=\mathcal{Q}E_p/\hbar$ with $E_p=E_0 \langle \hat{a}\rangle$, where $\mathcal{Q}$ is the dipole moment of transition $|2\rangle \leftrightarrow |3\rangle$. For simplicity, we assume the atoms are identically polarized so that the complex macroscopic polarization is ${\bf P}=\mathcal{NQ}\rho_{23}=\varepsilon_0 \chi_\text{p} E_\text{p}$, where $\mathcal{N}$ is the number density of atoms, $\rho$ is the density matrix and $\rho_{ij}=\langle i|\rho|j\rangle$, and $\chi_\text{p}$ denotes the susceptibility due to the atoms as seen by the probe field. Inducing a large $\chi_\text{p}$ requires $\mathcal{N}\lambda^3\gg 1$. In such dense dielectric medium the macroscopic probe field ${\bf E}_p$ that couples to atoms must be replaced by the local microscopic electric field ${\bf E}_\text{L}$, which is related to the macroscopic volume polarization ${\bf P}$ by the  Lorentz-Lorenz relation \cite{Jackson,*NDD1,*NDD2,UltraNTheo2}, namely ${\bf E}_L={\bf E}_p + {\bf P}/3\varepsilon_0$. This near dipole-dipole(NDD) effect leads to an effective susceptibility for ${\bf E}_\text{p}$, $\chi_\text{NDD}=\frac{\chi_\text{p}}{1-\chi_\text{p}/3}$ \cite{UltraNTheo2}.
The time evolution for the macroscopic density matrix is determined by
$\dot{\rho}  = -i (H\rho -\rho H) + \mathscr{L}\rho$, where $H = -\Delta_\text{p} \sigma_{22} - (\Delta_\text{p} + \Delta_\mu) \sigma_{11} + \Omega_\text{p} (\sigma_{32}+ \sigma_{23}) + \Omega_\mu (\sigma_{21}+ \sigma_{12})$ with operators $\sigma_{ij}=|i\rangle\langle j|$, and $\mathscr{L}\rho= \tilde{\gamma}/2\{2\hat{A}\rho\hat{A}^\dag - \hat{A}^\dag\hat{A} \rho - \rho \hat{A}^\dag\hat{A}\}$ with $\tilde{\gamma}=\{\gamma_1,\gamma_2,r\}$ corresponding to $\hat{A}=\{\sigma_{13},\sigma_{23},\sigma_{31}\}$.
 %
$\gamma_1 (\gamma_2)$ is the decay rate from $|3\rangle$ to $|1\rangle (|2\rangle)$. The decay rate can be estimated as $\gamma={4\omega_a^3 \mathcal{Q}^2 \sqrt{\varepsilon_h}}/{6\pi\varepsilon_0 \hbar c^3}$.
For simplicity, we assume $\gamma_1=\gamma_2=\gamma$ and neglect the decoherence of $|2\rangle$.
For a first-order analysis in the probe field ${\bf E}_p$ \cite{UltraNTheo1,UltraNTheo3}, we obtain the macroscopic polarization in the steady state
\begin{equation}\label{eq:chiNDD}\nonumber
 \chi_\text{p} = \frac{-s_0\gamma \left\{2i (r-2i\Delta_\mu)\left[ r+\gamma -2i (\Delta_p +\Delta_\mu)\right] - 8i \Omega_\mu^2 \right\} }{(r-2i\Delta_\mu) \left\{(\gamma -2i \Delta_p)\left[ r+\gamma -2i (\Delta_p +\Delta_\mu)\right] + 4\Omega_\mu^2\right\}} \;,
\end{equation}
where $s_0 ={\mathcal{N}\mathcal{Q}^2}/{\varepsilon_0 \hbar\gamma}= {3}\mathcal{N}\lambda^3/{8\pi^2\sqrt{\varepsilon_h}} \approx {\mathcal{N}\lambda^3}/{26\sqrt{\varepsilon_h}}$.

To obtain a large increase in the optomechanical coupling we need to accurately control the doping of the atoms so that $|\chi'_\text{p}/3-1|$ is small but much larger than $|\chi''_\text{p}|$,  which then yields $\chi_\text{NDD} \approx \chi'_\text{p}/(1-\chi'_\text{p}/3)\gg \chi'_\text{p}$.

Replacing $\chi$ with $\chi_\text{NDD}$ in Eqs. (\ref{eq:coeff}) and (\ref{eq:ratio}), we find that the coupling $g_\text{om,P}\propto\chi_\text{NDD}$, and thus a giant value of $|\chi_\text{NDD}|$ is obtained. The argument of the complex $\chi_\text{NDD}\in \mathbb{C}$, can be used to control the sign of the coupling, $g_\text{om}$. In Fig. \ref{fig:Dp}, we plot  $\chi_\text{NDD}$,  for both large and small mw driving, $\Omega_\mu/\gamma=\{1, 0.1\}$. In the case of large mw driving, and $\Delta_p \lesssim 0.303\gamma$, we find significant enhancement $|\chi'_\text{NDD}|>10^3$, while the medium is transparent or displays gain for the probe field ($\chi''_\text{NDD}\leq 0$). This induced gain can considerably reduce the intrinsic loss of the cavity and subsequently leads to $\kappa' \ll \kappa$ \cite{NarrowingGain1,NarrowingGain2,NarrowingGain3}. As a result, the CDR can be improved by more than three orders if $(\varepsilon_h-1)<\kappa/\kappa'$, see Eq. (\ref{eq:ratio}). 
 If a weaker mw drive is applied, e.g. $\Omega_\mu/\gamma=0.1$, (see Fig. \ref{fig:Dp}(b)), the atomic susceptibility drops precipitously  and we switch off the optomechanical coupling. Here $s_0=3/1.66$ requires a  number density  $\mathcal{N}=10^{14}-10^{15} ~\text{\centi\meter}^{-3}$, for $\lambda=0.5-1 ~\text{\micro\meter}$, corresponding to $\mathcal{N}\lambda^3\sim 100$.
\begin{figure}
  \centering
\setlength{\unitlength}{1cm}
\begin{picture}(9,2.5)
\put(0,-.4){ \includegraphics[width=.95\linewidth]{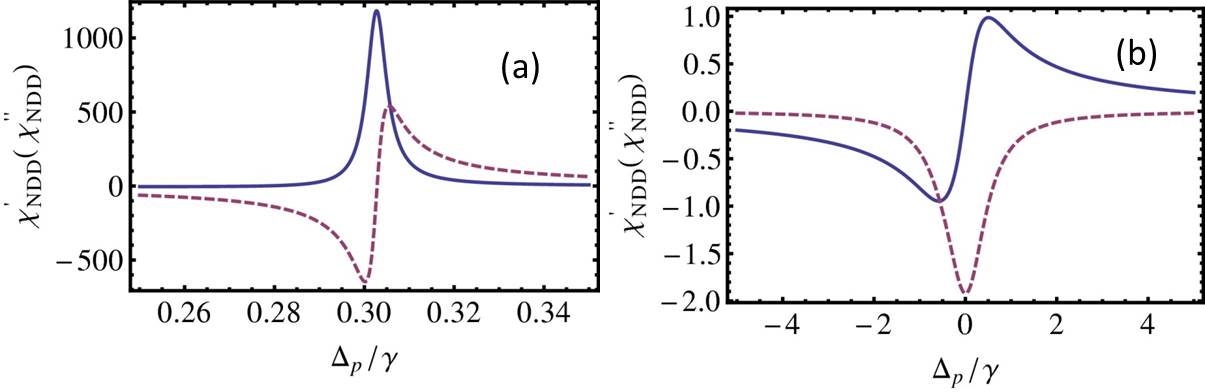}}
\end{picture}
 \caption{(Color online). Real (blue solid lines) and imaginary (red dashed lines) parts of the electric susceptibility $\chi_\text{NDD}$ as a function of the detuning $\Delta_p$ (a) $\Omega_\mu/\gamma=1$ and (b) $\Omega_\mu/\gamma=0.1$. $s_0=3/1.66, r=0.1\gamma, \Delta_\mu=0.4\gamma$ for $\chi_\text{NDD}$. $\chi''<0$ implies gain.}\label{fig:Dp}
\end{figure}

 We can now estimate the possible CDR that can be achieved by our scheme, $g_\text{om}/\kappa'$. Typically, in normal media $g_\text{om,h}/\kappa \sim 10^{-3}$ and $\varepsilon_h -1<10$, but due to the induced gain in our scheme, $\kappa/\kappa'(\varepsilon_h -1) >10$ is possible. Thus it is reasonable to assume $\frac{g_\text{om,h}}{\kappa' (\varepsilon_h-1)}=10^{-3}$ considering only the gain. When we fix the detuning, $\Delta_p=0.3\gamma, \Delta_\mu=0.4\gamma$, and the concentration of atoms, $s_0=3/1.66$, the SPSC regime, $g_\text{om}/\kappa'>1$, can be achieved over a large range of pump and mw drivings while the medium displays gain, see  Fig. \ref{fig:mwr}.
\begin{figure}
   \centering
\setlength{\unitlength}{1cm}
\begin{picture}(9,3)
\put(0,-.4){ \includegraphics[width=.95\linewidth]{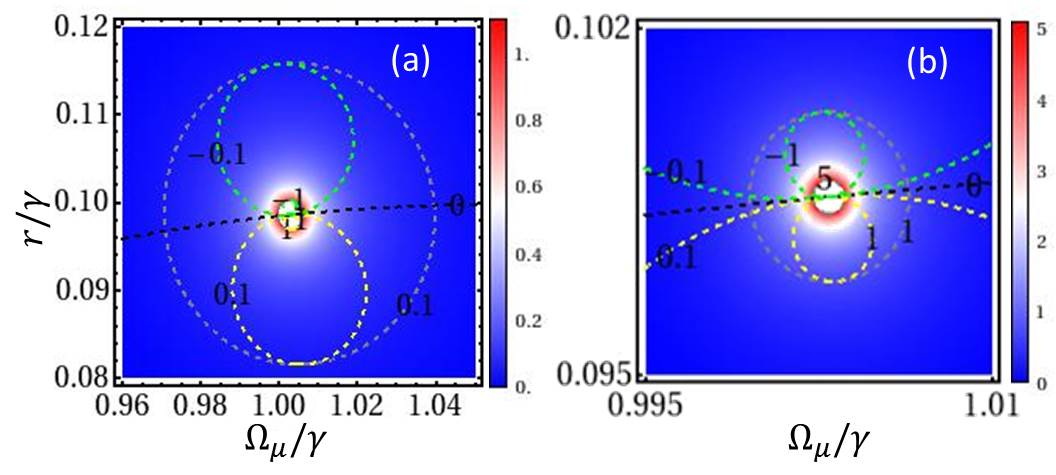}}
\end{picture}
 \caption{(Color online). Enhancement of optomechanical coupling-decay ratio (CDR): (a)  $|g_\text{om}/\kappa'|$ as a function of the pump rate $r$ and the mw field $\Omega_\mu$. Here $s_0=3/1.66,\;\; \Delta_p=0.3\gamma,\;\; \Delta_\mu=0.4\gamma$. (b) Enlarged central area of (a). Contours (dashed) are guide to the eye. In (b) (or (a)) dashed gray circle, overlapping with white spot, shows $|g_\text{om}/\kappa'|>1$ or $0.1$. Region above(below) the black line is gain(loss) and the yellow contour shows significant loss $10^{-3}\chi''> 0.1$ and gain,  green contour $10^{-3}\chi''< -0.1$, regions.}\label{fig:mwr}
\end{figure}
Normally, the density of atoms is fixed once the device is made and is therefore hard to control accurately. Figure \ref{fig:mwN}(a) shows that SPSC can be realized over a range of number densities $1.583<s_0<1.625$, by tuning the mw driving (overlap region between green and black contours). Interestingly, the phase of the coupling $g_\text{om}$, switches from $0\leftrightarrow \pi$, indicating a change in sign of the optomechanical coupling  when the mw driving is tuned across the white-blue boundary in  Fig. \ref{fig:mwN}(b), due a change in sign in  $\chi'_\text{NDD}$. Note that the direction of the radiation pressure, $F_\text{rp}= - \hbar g_\text{om}\langle \hat{a}^\dag \hat{a}\rangle/z_{zp}$ is dependent on the sign of the coupling \cite{OptoRev1}, and we find therefore, that the radiation pressure can be dynamically tuned from being repulsive to attractive. 
\begin{figure}
   \centering
\setlength{\unitlength}{1cm}
\begin{picture}(9,3.3)
\put(0,-.4){ \includegraphics[width=.95\linewidth]{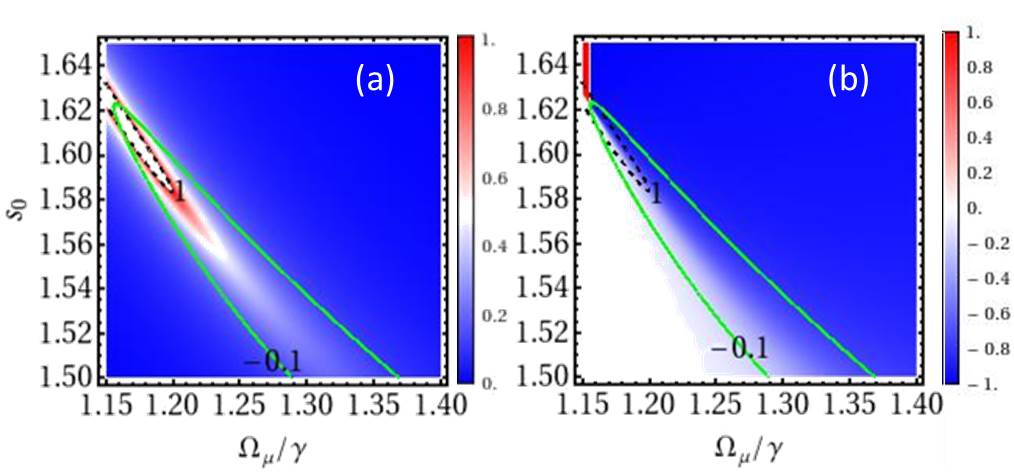}}
\end{picture}
 \caption{(Color online). Enhancement of the CDR over a range of concentration for the atomic dopant in the resonator. Modulus (a) and (b) argument [scaled by $\pi$],  of the ratio $g_\text{om}/\kappa'$ for different mw driving $r\Omega_\mu$ and number density $\mathcal{N}$. We choose $r=0.098\gamma,\, \Delta_p=0.3\gamma, \,\Delta_\mu=0.4\gamma$. Black contour shows the strong coupling region $g_\text{om}/\kappa'>1$, while green contour for a gain medium, $10^{-3}\chi''<-0.1$ (The gain inside the contour is larger).}\label{fig:mwN}
\end{figure}

Now we present possible implementations using  existing experimental systems. 
 Atomic $\Lambda$-type systems have been demonstrated in various solid-state systems from quantum dot \cite{CPTQD}, Nitrogen-vacancy (NV) \cite{CPTNV1,*CPTNV2,CPTNV3} or Silicon-vacancy (SiV) \cite{CPTSiV1,*CPTSiV2}, in diamond to atoms \cite{CPTAtom1,*CPTAtom2,*CPTAtom3,*CPTAtom4,*CPTAtom5}, and rare earth ions \cite{CPTREI1,*CPTREI2,*CPTREI3}. For our quantum configuration, we consider $\text{Er}^{3+}$  \cite{CPTREI1,*CPTREI2,*CPTREI3,NarrowingGain1,*NarrowingGain2} at low temperature ($T=10~\text{\kelvin}$), implanted in Silicon Nitride ($\text{Si}_3\text{N}_4$), membrane with $\varepsilon_h=4$ \cite{MembSiN1}. By choosing $s_0=3/1.66$ corresponding to $\mathcal{N}\approx 2.57\times 10^{13} ~\text{\centi\meter}^{-3}$, and $\Delta_p =0.30285\gamma$, at wavelength $\lambda=1550 ~\text{\nano\meter}$, we can induce a susceptibility of $\chi_\text{NDD}= 1181.45 -0.70 i$, at $\sin^2(kz_0)=0.5$, when $r/\gamma=0.1$, and $\Omega_\mu/\gamma=1$. For the cavity, we take $L=100\lambda, F=2\times 10^5$, yielding $Q_c=2\times 10^7~ (\kappa/2\pi=9.7 ~\text{\mega\hertz})$,, and $\kappa/\kappa' =30$ \cite{HFcavity}. For a $\text{Si}_3\text{N}_4$ membrane with thickness $l=100 ~\text{\nano\meter}$, diameter $D=10 ~\text{\micro\meter}$, mass density $\rho_m=2.7 ~\text{\gram}\cdot\text{cm}^{-3}$ ($m=21~\text{\pico\gram}$), a tensile stress of ${\rm T}_\text{s}\sim0.9 ~\text{\giga P}$ and $Q_m=4\times 10^6$, \cite{Membrane1,MembSiN1,NJPMembrane}, yields a motional oscillation frequency $\Omega_m/2\pi =40.8 ~\text{\mega\hertz}$, zero point motion $z_\text{zp}=3.1 ~\text{\femto\meter}$, and $\mathcal{B}=0.92$. From this we obtain a very large CDR, $g_\text{om}/\kappa' = 5.3$, and more importantly, an extremly large quantum cooperativity $\mathcal{C}_\text{Q}=g^2_\text{om}/\kappa' \gamma^*_\text{m}=174.7$ at the single-photon level, where $\gamma^*_\text{m}=\gamma_m\times \bar{n}_\text{th}$ is the mechanical decoherence rate  at $10~\text{\kelvin}$ ($\bar{n}_\text{th}=5.1\times 10^4$) \cite{QuantCoh,OptoRev2}.
If we take $\text{Cr}^{3+}$ ($\lambda=694~\text{\nano\meter}$) in Ruby ($l=50~\text{\nano\meter}, \rho_m=3.98 ~\text{\gram}\cdot\text{cm}^{-3}, T_s=0.3 ~\text{\giga\pascal}$) at $T=300 ~\text{\kelvin}$  \cite{CPTREI1,*CPTREI2} but assume lower finesse of $F=2\times 10^4$ yielding $Q_c=2\times 10^6$ as an estimation of absorption in Ruby \cite{AbsRuby1,*AbsRuby2}, we can still achieve $g_\text{om}/\kappa' = 22.4$ and $\mathcal{C}_\text{Q}\approx 230.8$ at {\em room temperature}.
As mentioned above, $\mathcal{C}_\text{Q}>1$ allows a number of coherent optical control techniques to manipulate the quantum state of the mechanics.

In conclusion, based on our coupled mode analysis a giant enhancement of the optomechanical coupling can be achieved by inducing an ultrarefractive index in the membrane in the middle based optomechanical system. This giant enhancement permits low temperature explorations of the fascinating regimes of SPSC and QC in excess of unity and for certain materials, e.g. Ruby, a {\em room temperature implementation} may be possible. By tuning the pump laser or mw driving we can rapidly switch on and off the optomechanical coupling or change between repulsive and attractive optomechanical forces.


\FloatBarrier
%

\end{document}